\documentclass[preprint,showpacs,preprintnumbers,amsmath,amssymb]{revtex4}

\pdfoutput=1

\usepackage{graphicx}
\usepackage{dcolumn}
\usepackage{bm}
\usepackage{amssymb}

\begin{document}
\title{
Quantum tunneling and evolution speed
in an exactly solvable coupled double-well system
}

\author{Hideo Hasegawa}
\altaffiliation{hideohasegawa@goo.jp}
\affiliation{Department of Physics, Tokyo Gakugei University,  
Koganei, Tokyo 184-8501, Japan}%

\date{\today}

\begin{abstract}
Exact analytical calculations of eigenvalues and eigenstates are presented
for quantum coupled double-well (DW) systems with Razavy's hyperbolic potential.
With the use of four kinds of initial wavepackets, we have calculated the tunneling period $T$ 
and the orthogonality time $\tau$ which signifies 
a time interval for an initial state to evolve to its orthogonal state.
We discuss the coupling dependence of $T$ and $\tau$, and 
the relation between $\tau$ and the concurrence $C$ 
which is a typical measure of the entanglement in two qubits.
Our calculations have shown that it is not clear whether the speed of quantum evolution
may be measured by $T$ or $\tau$ and that the evolution speed measured by $\tau$ (or $T$) 
is not necessarily increased with increasing $C$.
This is in contrast with the earlier study
[V. Giovannetti, S. Lloyd and L. Maccone, Europhys. Lett. {\bf 62} (2003) 615]
which pointed out that the evolution speed measured by $\tau$ is enhanced by the entanglement
in the two-level model.

\vspace{0.5cm}

\end{abstract}
        

\maketitle
\newpage
\section{Introduction}
Double-well (DW) potential models have been extensively employed in various fields
of quantum physics, in which the tunneling is one of fascinating quantum effects.
Although quartic DW potentials are commonly adopted for the theoretical study,
one cannot obtain their exact eigenvalues and eigenfunctions
of the Schr\"{o}dinger equation.
Then it is necessary to apply various approximate approaches
such as perturbation and spectral methods to quartic potential models \cite{Tannor07}.
Razavy \cite{Razavy80} proposed the quasi-exactly solvable hyperbolic DW potential, 
for which one may exactly determine a part of whole eigenvalues and eigenfunctions.
A family of quasi-exactly solvable potentials has been investigated
\cite{Finkel99,Bagchi03}.

The two-level (TL) system which is a simplified model of a DW system, 
has been employed for a study on qubits which play important roles 
in quantum information and quantum computation \cite{Ref1}.
The relation between the entanglement and the speed of evolution
of TL systems has been discussed \cite{Giovannetti03,Giovannetti03b,Batle05,Curilef06}.
The entanglement in qubits has been studied with the use of uncoupled and coupled 
TL models \cite{Batle05,Curilef06,Zander13}.
In recent years, several experimental studies for coupled TL systems 
have been reported \cite{Pashkin03,Majer05}.

In contrast to the simplified TL model mentioned above, studies 
on {\it coupled} DW systems are scanty, as far as we are aware of.
This is because a calculation of a coupled DW system is much tedious than 
those of a single DW system and of a coupled TL model. 
In the present study, we adopt coupled two DW systems, each of which is described by Razavy's potential. 
One of advantages of our model is that we may exactly determine eigenvalues 
and eigenfunctions of the coupled DW system. We study the tunneling period $T$ and
the orthogonality time $\tau$ which stands for the time interval 
for an assumed initial state to develop to its orthogonal state.
We investigate the relation between the speed of quantum evolution measured by $\tau$ and 
the concurrence which is one of typical measures of entanglement of two qubits.
The difference and similarity between results in our coupled DW system and 
the TL model \cite{Giovannetti03,Giovannetti03b,Batle05,Curilef06} are discussed.
These are purposes of the present paper.

The paper is organized as follows.
In Sec. II, we describe the calculation method employed in our study,
briefly explaining Razavy's potential \cite{Razavy80}.
Exact analytic expressions for eigenvalues and eigenfunctions for
coupled DW systems are presented. In Sec. III,
with the use of four kinds of initial wavepackets, we perform model calculations
of tunneling period $T$ and the orthogonality time $\tau$.
In Sec. IV, we discuss the relation between the calculated $\tau$ and the concurrence.
Sec. V is devoted to our conclusion.

\section{The adopted method}
\subsection{Coupled double-well system with Razavy's potential}
We consider coupled two DW systems whose Hamiltonian is given by 
\begin{eqnarray}
H &=& \sum_{n=1}^2 \left[ -\frac{\hbar^2}{2m} \frac{\partial^2}{\partial x_n^2} 
+ V(x_n )\right] - g x_1 x_2, 
%
\label{eq:H1}
\end{eqnarray}
with
\begin{eqnarray}
V(x) &=& \frac{\hbar^2}{2m}
\left[\frac{\xi^2}{8} \:{\rm cosh} \:4x - 4 \xi \:{\rm cosh} \:2x- \frac{\xi^2}{8}
\right],
\label{eq:H3}
\end{eqnarray}
where $x_1$ and $x_2$ stand for coordinates of two distinguishable particles 
of mass $m$ in double-well systems coupled by an interaction $g$, and
$V(x)$ signifies Razavy's potential \cite{Razavy80}.
The potential $V(x)$ with $\hbar=m=\xi=1.0$ adopted in this study
is plotted in Fig. 1(a).
Minima of $V(x)$ locate at $x_s=\pm 1.38433$ with $V(x_s)=-8.125$
and its maximum is $V(0)=-2.0$ at $x=0.0$.

First we consider the case of $g=0.0$ in Eq. (\ref{eq:H1}).
Eigenvalues of Razavy's double-well potential of Eq. (\ref{eq:H3}) are given by \cite{Razavy80}
\begin{eqnarray}
\epsilon_0 &=& \frac{\hbar^2}{2m}\left[ -\xi -5 -2 \sqrt{4-2 \xi+\xi^2} \right], \\
\epsilon_1 &=& \frac{\hbar^2}{2m}\left[ \xi-5 -2 \sqrt{4+2 \xi+\xi^2} \right], \\
\epsilon_2 &=& \frac{\hbar^2}{2m}\left[ -\xi-5 +2 \sqrt{4-2 \xi+\xi^2} \right], \\
\epsilon_3 &=& \frac{\hbar^2}{2m}\left[ \xi-5 +2 \sqrt{4+2 \xi+\xi^2} \right]. 
\end{eqnarray}
Eigenvalues for the adopted parameters are $\epsilon_0=-4.73205$, $\epsilon_1=-4.64575$,
$\epsilon_2=-1.26795$ and  $\epsilon_3=0.645751$.
Both $\epsilon_0$ and $\epsilon_1$ locate below $V(0)$ as shown by dashed curves in Fig. 1(a),
and $\epsilon_2$ and $\epsilon_3$ are far above $\epsilon_1$. In this study,
we take into account the lowest two states of $\epsilon_0$ and $\epsilon_1$ 
whose eigenfunctions are given by \cite{Razavy80}
\begin{eqnarray}
\phi_0(x) &=& A_0 \; e^{-\xi \:{\rm cosh} \:2x/4} \left[3 \xi \:{\rm cosh} \:x
+(4-\xi+2 \sqrt{4-2 \xi+\xi^2})\: {\rm cosh}\: 3x \right], \\
\phi_1(x) &=&  A_1 \;e^{-\xi \:{\rm cosh} \:2x/4} \left[3 \xi \:{\rm sinh}\: x
+(4+\xi+2 \sqrt{4+2 \xi+\xi^2})\: {\rm sinh} \:3x \right],
%
%
\end{eqnarray}
$A_{n}$ ($n=0,1$) denoting normalization factors.
Figure 1(b) shows the eigenfunctions of $\phi_0(x)$ and $\phi_1(x)$, which 
are symmetric and anti-symmetric, respectively, with respect to the origin.

\begin{figure}
\begin{center}
\includegraphics[keepaspectratio=true,width=120mm]{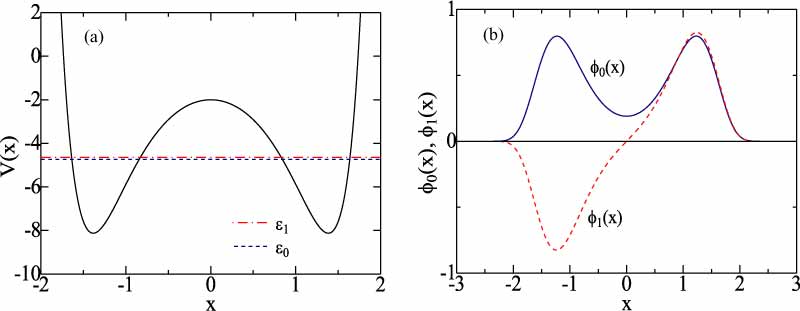}
\end{center}
\caption{
(Color online) 
(a) Razavy's DW potential $V(x)$ (solid curve),
dashed and chain curves expressing eigenvalues of $\epsilon_0$ and $\epsilon_1$,
respectively, for $\hbar=m=\xi=1.0$ [Eq.(\ref{eq:H3})].
(b) Eigenfunctions of $\phi_0(x)$ (solid curve) and $\phi_1(x)$ (dashed curve).
}
\label{fig1}
\end{figure}

\begin{figure}
\begin{center}
\includegraphics[keepaspectratio=true,width=150mm]{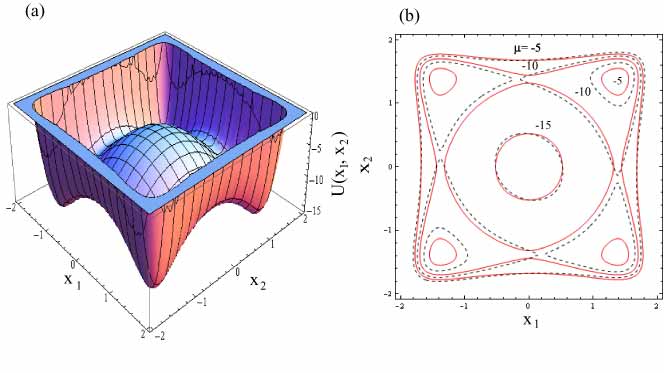}
\end{center}
\caption{
(Color online) 
(a) 3D plot of a composite potential $U(x_1, x_2)$ as functions of $x_1$ and $x_2$.
(b) Contour plots of $U(x_1, x_2)=\mu$ for $\mu=-15$, $-10$ and $-5$
with $g=0.0$ (solid curves) and $g=1.0$ (dashed curves).
}
\label{fig2}
\end{figure}

Figure \ref{fig2}(a) shows the 3D plot of the composite potential $U(x_1, x_2)$ defined by
\begin{eqnarray}
U(x_1, x_2) &=& V(x_1)+V(x_2) -g x_1 x_2.
\label{eq:H2} 
\end{eqnarray}
It has four minima of $U(\pm x_s, \pm x_s)=-16.25$
and one maximum of $U(0.0, 0.0)=-4.0$ for $g=0.0$.
Solid curves in Fig. \ref{fig2}(b) show contour plots of $U(x_1, x_2)=\mu$ for $\mu=-15$, -10 and -5
with $g=0.0$. For a comparison, dashed curves shows the result with $g=1.0$,
for which $U(\pm x_s, \mp x_s)-U(\pm x_s, \pm x_s)=3.8327$. 
Dashed curves with $g=1.0$ are slightly different from solid curves with $g=0.0$.

\subsection{Eigenvalues and eigenstates of the coupled DW system}
We calculate exact eigenvalues and eigenstates of the coupled two DW systems described 
by Eq. (\ref{eq:H1}). With basis states of 
$\phi_0 \phi_0$, $\phi_0 \phi_1$, $\phi_1 \phi_0$ and $\phi_1 \phi_1$
where $ \phi_n \phi_k \equiv  \phi_n(x_1) \phi_k(x_2)$,
the energy matrix for the Hamiltonian given by Eq. (\ref{eq:H1}) is expressed by
\begin{eqnarray}
{\cal H} &=& \left( {\begin{array}{*{20}c}
   {2 \epsilon_0 } & {0 } & {0 } & {-g \gamma^2} \\
   {0 } & {\epsilon_0 + \epsilon_1 } & {-g \gamma^2 } & {0} \\
   {0 } & {-g \gamma^2 } & {\epsilon_0 + \epsilon_1 } & {0} \\
   {-g \gamma^2 } & {0 } & {0 } & {2 \epsilon_1} \\   
\end{array}} \right),
\label{eq:H4}
\end{eqnarray}
with
\begin{eqnarray}
\gamma &=& \int_{-\infty}^{\infty} \phi_0(x)\: x \: \phi_1(x)\:dx=1.13823.
\label{eq:H5}
\end{eqnarray}
Eigenvalues of the energy matrix are given by
\begin{eqnarray}
E_0 &=& \epsilon -\sqrt{\delta^2+ g^2 \gamma^4}, 
\label{eq:H6a}\\
E_1 &=& \epsilon - g \gamma^2, \\
E_2 &=& \epsilon + g \gamma^2, \\
E_3 &=& \epsilon + \sqrt{\delta^2+ g^2 \gamma^4},
\label{eq:H6b}
\end{eqnarray}
where
\begin{eqnarray}
\epsilon &=& \epsilon_1+\epsilon_0=-9.3778, \\
\delta &=& \epsilon_1-\epsilon_0=0.0863.
\end{eqnarray}
Corresponding eigenfunctions are given by
\begin{eqnarray}
\Phi_0(x_1,x_2) &=& \cos \theta \:\phi_0(x_1) \phi_0(x_2) 
+ \sin \theta \:\phi_1(x_1) \phi_1(x_2), 
\label{eq:H7a}\\
\Phi_1(x_1,x_2) &=& \frac{1}{\sqrt{2}} \left[ \phi_0(x_1) \phi_1(x_2)
+ \phi_1(x_1) \phi_0(x_2) \right], \\
\Phi_2(x_1,x_2) &=& \frac{1}{\sqrt{2}} \left[- \phi_0(x_1) \phi_1(x_2)
+ \phi_1(x_1) \phi_0(x_2) \right], \\
\Phi_3(x_1,x_2) &=& -\sin \theta \:\phi_0(x_1) \phi_0(x_2)
+ \cos \theta \: \phi_1(x_1) \phi_1(x_2),
\label{eq:H7b}
\end{eqnarray}
where
\begin{eqnarray}
\tan \:2 \theta &=& \frac{g \gamma^2}{\delta}.
\;\;\;\;\mbox{$\left(-\frac{\pi}{4} \leq \theta \leq \frac{\pi}{4} \right)$}
\label{eq:H7c}
\end{eqnarray}
Eigenvalues $E_{\nu}$ ($\nu=0$-3) are plotted as a function of $g$ in Fig. 3, which is
symmetric with respect to $g=0.0$. 
For $g=0.0$, $E_1$ and $E_2$ are degenerate. We hereafter study the case of $g \geq 0.0$.
With increasing $g$, energy gaps between $\epsilon_0$ and $\epsilon_1$ and between $\epsilon_2$ 
and $\epsilon_3$ are gradually decreased while that between $\epsilon_1$ and $\epsilon_2$ is increased.

\begin{figure}
\begin{center}
\includegraphics[keepaspectratio=true,width=70mm]{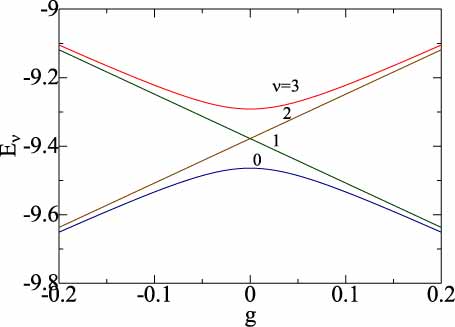}
\end{center}
\caption{
(Color online) 
Eigenvalues $E_{\nu}$ ($\nu=0-$3) of a coupled DW system as a function 
the coupling strength $g$.
}
\label{fig3}
\end{figure}

The time-dependent wavepacket is expressed by
\begin{eqnarray}
\Psi(t) &=& \Psi(x_1,x_2,t)
= \sum_{\nu=0}^{3}\:a_{\nu} \: \Phi_{\nu}(x_1,x_2) \:e^{-i E_{\nu} t/\hbar},
\label{eq:G1}
\end{eqnarray}
where expansion coefficients $a_{\nu}$ satisfy the relation
\begin{eqnarray}
\sum_{\nu=0}^3 \vert a_{\nu} \vert^2 &=& 1.
\end{eqnarray}
Expansion coefficients $a_{\nu}$ may be formally determined for a given initial wavepacket,
which requires cumbersome calculations. In this study they are 
assumed {\it a priori} as will be given shortly.

The correlation function $\Gamma(t)$ is defined by
\begin{eqnarray}
\Gamma(t) &=& \vert \int_{-\infty}^{\infty} \int_{-\infty}^{\infty} 
\Psi^*(x_1,x_2, 0) \:\Psi(x_1,x_2,t)\;dx_1\:dx_2 \: \vert, \\
&=& \vert \; \vert a_0 \vert^2  + \sum_{\nu=1}^3 \: \vert a_{\nu} \vert^2 
\:e^{- i \Omega_{\nu} t} \: \vert,
\label{eq:G2}
\end{eqnarray}
where $\Omega_{\nu}=(E_{\nu}-E_0)/\hbar$.
The tunneling period $T$ for the initial wavepacket given by Eq. (\ref{eq:G1}) is determined by
\begin{eqnarray}
T &=& \min_{\forall \:t \:> 0}\; \{ \Gamma(t)=1 \}.
\label{eq:E1}
\end{eqnarray} 
On the contrary, the orthogonality time $\tau$ is 
provided by the time interval such that an initial wavepacket takes to evolve into
the orthogonal state \cite{Giovannetti03,Giovannetti03b,Batle05,Curilef06},
\begin{eqnarray}
\tau &=& \min_{\forall \:t \:> 0}\; \{ \Gamma(t)=0 \}.
\label{eq:E2}
\end{eqnarray}

In the case of a simple wavepacket including only two states, {\it e.g.}
$a_{\nu}=(1/\sqrt{2}) \:(\delta_{\nu, 0}+\delta_{\nu, \kappa})$, the correlation function becomes
\begin{eqnarray}
\Gamma(t) &=& \frac{1}{2} \vert 1+ e^{-i \Omega_{\kappa} t} \vert
= \sqrt{ \frac{1+ \cos \Omega_{\kappa} t}{2} },
\label{eq:E3}
\end{eqnarray}
for which we easily obtain $T$ and $\tau$
\begin{eqnarray}
T &=& 2 \tau= \frac{2 \pi}{\Omega_{\kappa}}.
\label{eq:E4}
\end{eqnarray}

In the case of $g=0.0$ where $\Omega_1=\Omega_2=\Omega_3/2$, 
Eqs. (\ref{eq:E1}) and (\ref{eq:E2}) become
\begin{eqnarray}
T &=& \min_{\forall \:t \:> 0}\; \large\{\vert a_0 \vert^2 
+ (\vert a_1 \vert^2+ \vert a_2 \vert^2)\:z(t) 
+ \vert a_3 \vert^2 \:z(t)^2=1 \large\}, \\
\tau &=& \min_{\forall \:t}\; \large\{\vert a_0 \vert^2 
+ (\vert a_1 \vert^2+ \vert a_2 \vert^2)\:z(t) + \vert a_3 \vert ^2 z(t)^2 =0 \large\}, 
\end{eqnarray}
where $z(t)=e^{- i \Omega_1 t}$.
Solutions of $T$ and $\tau$ may be obtainable from roots of respective 
polynomial equations for $z(t)$ \cite{Batle05,Curilef06}.

In a general case, however, $T$ and $\tau$ are obtained 
by solving Eqs. (\ref{eq:E1}) and (\ref{eq:E2}) with a numerical method, as will be shown later.

\section{Model calculations}
Dynamical properties of wavepackets A-D with assumed expansion coefficients shown in Table 1, 
have been studied. We will report results of the case with $g=0.0$ for wavepackets A and B
in Sec. III A, and those with $g=0.1$ for wavepackets C and D in Sec. III B.

\begin{center}
\begin{tabular}[b]{|c|c|c|c|c|c| }
\hline
wavepacket & $a_0$ & $a_1$ &  $ a_2$ & $ a_3 $  \\ 
\hline \hline
$\;\;\;$ A [Eq. (\ref{eq:A2})] $\;\;\;$ &  $\;\;\;\; \frac{1}{2} \;\;\;\;$  &
  $\;\; \frac{1}{\sqrt{2}} \;\;$ 
& $\;\;\; 0 \;\;\;$ & $\;\;\; \frac{1}{2} \;\;\;$  \\
\hline
B [Eq. (\ref{eq:B2})]   &  $\frac{1}{\sqrt{2}}$  &  0  & 0 & $\frac{1}{\sqrt{2}}$   \\
\hline
C [Eq. (\ref{eq:C1})] &  $\frac{1}{\sqrt{2}} $ &  $\frac{1}{\sqrt{2}}$ & 0 &  0 \\
\hline
D [Eq. (\ref{eq:D1})]  &  $\frac{1}{2}$ &  $\;\;\frac{1}{2} \;\;$ & 
$\;\;\frac{1}{2}\;\;$ & $\frac{1}{2}$ \\
\hline
\end{tabular}
\end{center}
{\it Table 1} Assumed expansion coefficients $a_{\nu}$ ($\nu=0$ to 3) for
four wavepackets A, B, C and D. 

\subsection{Uncoupled double-well system ($g=0.0$)}
First we consider the uncoupled DW with $g=0.0$, for which eigenvalues are
\begin{eqnarray}
E_0 &=& -9.4641, \;\;\;
E_1 = E_2= -9.3778, \;\;\;
E_3 = -9.2915,
\end{eqnarray}
leading to
\begin{eqnarray}
\Omega_1 &=& \Omega_2=0.0863, \;\;\;\Omega_3=0.1726,
\end{eqnarray}
and eigenfunctions are given by
\begin{eqnarray}
\Phi_0(x_1,x_2) &=& \phi_0(x_1) \phi_0(x_2), \\
\Phi_1(x_1,x_2) &=& \frac{1}{\sqrt{2}} \left[ \phi_0(x_1) \phi_1(x_2)
+ \phi_1(x_1) \phi_0(x_2) \right], \\
\Phi_2(x_1,x_2) &=& \frac{1}{\sqrt{2}} \left[- \phi_0(x_1) \phi_1(x_2)
+ \phi_1(x_1) \phi_0(x_2) \right], \\
\Phi_3(x_1,x_2) &=& \phi_1(x_1) \phi_1(x_2).
\end{eqnarray}
Figure 4(a), 4(b), 4(c) and 4(d) show eigenfunctions $\Phi_{\nu}(x_1,x_2)$
for $\nu=0$, 1, 2 and 3, respectively.

\begin{figure}
\begin{center}
\includegraphics[keepaspectratio=true,width=120mm]{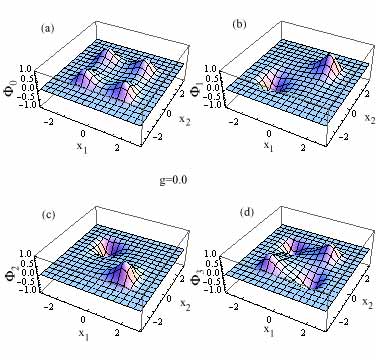}
\end{center}
\caption{
(Color online) 
Eigenfunctions of (a) $\Phi_0(x_1,x_2)$, (b) $\Phi_1(x_1,x_2)$, (c) $\Phi_2(x_1,x_2)$,
and (d) $\Phi_3(x_1,x_2)$ for $g=0.0$.
}
\label{fig4}
\end{figure}

\subsubsection{Wavepacket A: $a_0=1/2$, $a_1=1/\sqrt{2}$, $a_2=0$ and $a_3=1/2$}
\begin{figure}
\begin{center}
\includegraphics[keepaspectratio=true,width=80mm]{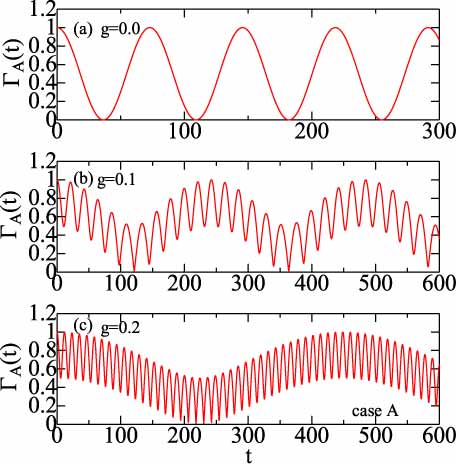}
\end{center}
\caption{
(Color online) 
Correlation functions $\Gamma_A(t)$ with (a) $g=0.0$, (b) $g=0.1$ and (c) $g=0.2$ 
for the wavepacket A.
}
\label{fig5}
\end{figure}

A factorizable product state is expressed by 
\begin{eqnarray}
\Psi_{prod}&=& \Psi_{RR}(x_1,x_2) = \Psi_R(x_1) \Psi_R(x_2), 
\label{eq:A1a}\\
&=& \frac{1}{2} \left[ \phi_0(x_1) \phi_0(x_2)+ \phi_0(x_1) \phi_1(x_2) 
+\phi_1(x_1) \phi_0(x_2)+\phi_1(x_1) \phi_1 (x_2)\right], \\
&=& \frac{1}{2} \left[ \Phi_0(x_1,x_2) +\Phi_3(x_1,x_2)  \right]
+ \frac{1}{\sqrt{2}} \Phi_1(x_1,x_2),
\label{eq:A1}
\end{eqnarray}
where magnitude of $\Psi_R(x_{\nu})$ $( =[\phi_0(x_{\nu})+\phi_1(x_{\nu})]/\sqrt{2} )$ 
localizes at the right well in the $x_{\nu}$ axis ($\nu=1, \:2$).
The wavepacket yielding initially the product state given by Eq. (\ref{eq:A1})
is described by
\begin{eqnarray}
\Psi_{A}(x_1,x_2,t) &=& \frac{1}{2} 
\left[ \Phi_0(x_1,x_2)\:e^{-i E_0 t/\hbar} +\Phi_3(x_1,x_2)\:e^{-i E_3 t/\hbar}  \right]
+ \frac{1}{\sqrt{2}} \Phi_1(x_1,x_2)\:e^{-i E_1 t/\hbar},
\label{eq:A2}
\end{eqnarray}
and the relevant correlation function is given by
\begin{eqnarray}
\Gamma_A(t) &=& \vert \frac{1}{4}\:\left( 1+ e^{-i \Omega_3 t/\hbar }\right)
+\frac{1}{2}\: e^{-i \Omega_1 t} \vert,
\label{eq:A3}
\end{eqnarray}
where $\Omega_1=\Omega_3/2=0.0863$.
Calculated $\Gamma_A(t)$ is plotted in Fig. \ref{fig5}(a) which
yields the tunneling period of $T=2 \pi/\Omega_1=72.81$
and the orthogonality time of $\tau=T/2=36.40$.
Figures \ref{fig5}(b) and \ref{fig5}(c) will be explained later (Sec. IV B).

\begin{figure}
\begin{center}
\includegraphics[keepaspectratio=true,width=150mm]{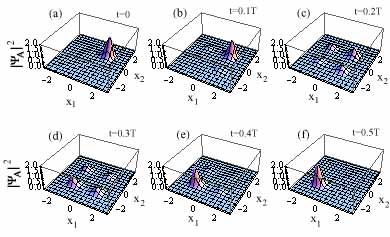}
\end{center}
\caption{
(Color online) 
Time-dependent magnitudes of $\vert \Psi_A(x_1,x_2,t) \vert^2$ 
at (a) $t=0.0$, (b) $t=0.1 T$, (c) $t=0.2 T$, (d) $t=0.3 T$, (e) $t=0.4 T$ and (f) $t= 0.5T$
for the wavepacket A [Eq.(\ref{eq:A2})] with $g=0.0$ where $T=72.81$.
Magnitudes of wavepackets at $t=0.6T$, $0.7T$, $0.8T$, $0.9T$ and $T$ are
the same as those at $t=0.4T$, $0.3T$, $0.2T$, $0.1T$ and $0$, respectively.
}
\label{fig6}
\end{figure}

Time-dependent magnitudes of $\vert \Psi_A(x_1, x_2, t) \vert^2$ are 
shown in Figs. \ref{fig6}(a)-\ref{fig6}(f). Figure \ref{fig6}(a) shows that
the wavepacket initially  has the maximum magnitude at the $RR$ side of
$(x_1,x_2)=(x_m, x_m)$ with $x_m=1.23534$
near the bottom of the right-side well of $U(x_s, x_s)$ with $x_s=1.38433$,
where $RR$ signifies the right side in the $x_1$ axis and the right side in $x_2$ axis.
At $t=0.2 T$, $\vert \Psi_A(x_1, x_2, t) \vert^2$ in Fig \ref{fig6}(b)
has finite magnitudes near $LL$, $RL$ and $LR$ sides besides $RR$ one.
This implies a tunneling of particles among four bottoms of $U(\pm x_s, \pm x_s)$. 
The orthogonal state to Eq. (\ref{eq:A1}) is given by
\begin{eqnarray}
\Psi_{LL}(x_1,x_2) &=& \Psi_L(x_1) \Psi_L(x_2), \\
&=& \frac{1}{2} \left[ \phi_0(x_1) \phi_0(x_2)- \phi_0(x_1) \phi_1(x_2) 
-\phi_1(x_1) \phi_0(x_2)+\phi_1(x_1) \phi_1 (x_2)\right], \\
&=& \frac{1}{2} \left[ \Phi_0(x_1,x_2) +\Phi_3(x_1,x_2)  \right]
- \frac{1}{\sqrt{2}} \Phi_1(x_1,x_2),
\end{eqnarray}
where magnitude of $\Psi_L(x_{\nu})$ $( =[\phi_0(x_{\nu})-\phi_1(x_{\nu})]/\sqrt{2} )$ 
localizes at the left well in the $x_{\nu}$ axis ($\nu=1, 2$).
$\Psi_A(x_1, x_2, t)$ reduces to $\Psi_{LL}(x_1, x_2)$ at $t=0.5T$,
and it returns to $\Psi_{RR}(x_1, x_2)$ at $t=T$.

\begin{figure}
\begin{center}
\includegraphics[keepaspectratio=true,width=150mm]{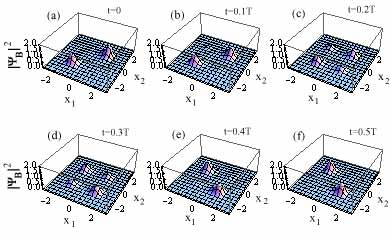}
\end{center}
\caption{
(Color online) 
Time-dependent magnitudes of $\vert \Psi_B(x_1,x_2,t) \vert^2$ 
at (a) $t=0.0$, (b) $t=0.1 T$, (c) $t=0.2 T$, (d) $t=0.3 T$, (e) $t=0.4 T$ and (f) $t=0.5 T$
for the wavepacket B [Eq.(\ref{eq:B2})] with $g=0.0$ where $T=36.40$.
}
\label{fig7}
\end{figure}

\subsubsection{Wavepacket B: $a_0=1/\sqrt{2}$, $a_1=a_2=0.0$ and $a_3=1/\sqrt{2}$}
As a typical entangled state which cannot be expressed in a factorized form, 
we consider the state
\begin{eqnarray}
\Psi_{ent}(x_1,x_2) &=& 
\frac{1}{\sqrt{2}} \left[ \phi_0(x_1) \phi_0(x_2)+\phi_1(x_1) \phi_1(x_2)  \right], \\
&=& \frac{1}{\sqrt{2}} \left[ \Phi_0(x_1,x_2) +\Phi_3(x_1,x_2)  \right].
\label{eq:B1}
\end{eqnarray}
The relevant wavepacket is expressed by
\begin{eqnarray}
\Psi_{B}(x_1. x_2, t) &=& \frac{1}{\sqrt{2}} \left[ \Phi_0(x_1,x_2)\:e^{-i E_0 t/\hbar} 
+\Phi_3(x_1,x_2)\:e^{-i E_3 t/\hbar}  \right],
\label{eq:B2}
\end{eqnarray}
and its correlation function is given by
\begin{eqnarray}
\Gamma_B(t) &=& \frac{1}{2}\vert 1+ e^{-i \Omega_3 t} \vert
= \sqrt{ \frac{1+\cos \Omega_3 t}{2} },
\label{eq:B3}
\end{eqnarray}
where $\Omega_3=0.1726$.
The tunneling period becomes $T=2 \pi/\Omega_3=36.40$
and the orthogonality time is given by $\tau=T/2=18.20$.

Figures \ref{fig7}(a)-\ref{fig7}(f) show the time-dependent magnitudes
of $\vert \Psi_{B}(x_1,x_2) \vert^2$ at $0 \leq t \leq T$.
Initially $\vert \Psi_{B}(x_1,x_2) \vert^2$ has peaks at both $RR$ and $LL$ sides.
At $t=0.5 T$, it reduces to
\begin{eqnarray}
\Psi_{ent}^{\bot}(x_1,x_2) &=& 
\frac{1}{\sqrt{2}} \left[ \Phi_0(x_1,x_2) - \Phi_3(x_1,x_2)  \right],
\label{eq:B3b}
\end{eqnarray}
which is orthogonal to the assumed initial state given by Eq. (\ref{eq:B1}) and
which has peaks at both $RL$ and $LR$ sides.

\subsection{Coupled double-well system ($g=0.1$)}
\begin{figure}
\begin{center}
\includegraphics[keepaspectratio=true,width=120mm]{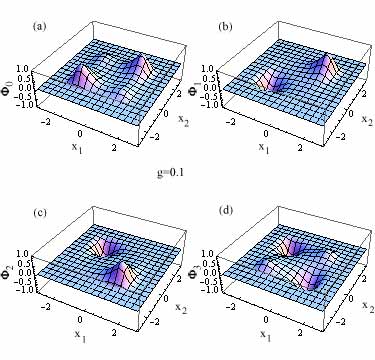}
\end{center}
\caption{
(Color online) 
Eigenfunctions of (a) $\Phi_0(x_1,x_2)$, (b) $\Phi_1(x_1,x_2)$, (c) $\Phi_2(x_1,x_2)$,
and (d) $\Phi_3(x_1,x_2)$ for $g=0.1$.
}
\label{fig8}
\end{figure}

Next we study coupled DW systems with $g=0.1$, for which eigenvalues are
\begin{eqnarray}
E_0 &=& -9.53347, \;\; E_1=-9.50736,\;\;  E_2=-9.24825,\;\;  E_3=-9.22213, 
\end{eqnarray}
leading to
\begin{eqnarray}
\Omega_1&=& 0.02611,  \;\; \Omega_2=0.28522, \;\;\Omega_3=0.31134.
\end{eqnarray}
The potential difference between the two bottoms 
is $U(\pm x_s, \mp x_s)-U(\pm x_s, \pm x_s)=0.38327$.
Figures \ref{fig8}(a)-\ref{fig8}(d) show eigenfunctions $\Phi_{\nu}(x_1,x_2)$ for $\nu=0-3$. 

\subsubsection{Wavepacket C: $a_0=a_1=1/\sqrt{2}$ and $a_2=a_3=0$}

With $a_0=a_1=1/\sqrt{2}$ and $a_2=a_3=0$, the wavepacket in Eq. (\ref{eq:G1}) becomes
\begin{eqnarray}
\Psi_C(x_1,x_2,t) &=& \frac{1}{\sqrt{2}} 
\left[ \Phi_0(x_1,x_2) \:e^{-i E_0 t/\hbar}+\Phi_1(x_1,x_2) \:e^{-i E_1 t/\hbar} \right],
\label{eq:C1}
\end{eqnarray}
whose correlation function is given by
\begin{eqnarray}
\Gamma_C(t) &=& \frac{1}{2} \vert 1 + e^{-i \Omega_1 t} \vert
= \sqrt{ \frac{1+ \cos \Omega_1 t}{2} },
\label{eq:C2}
\end{eqnarray}
with $\Omega_1=0.02611$. The tunneling period is $T=2 \pi/\Omega_1=240.63$
and the orthogonality time is $\tau=T/2=120.32$.

\begin{figure}
\begin{center}
\includegraphics[keepaspectratio=true,width=150mm]{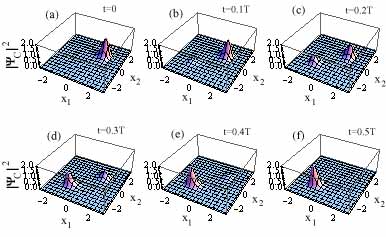}
\end{center}
\caption{
(Color online) 
Time-dependent magnitudes of $\vert \Psi_C(x_1,x_2,t) \vert^2$ 
at (a) $t=0.0$, (b) $t=0.1 T$, (c) $t=0.2 T$, (d) $t=0.3 T$, (e) $t=0.4 T$ and (f) $t= 0.5 T$
for the wavepacket C [Eq. (\ref{eq:C1})] with $g=0.1$ where $T=240.63$.
}
\label{fig9}
\end{figure}

\begin{figure}
\begin{center}
\includegraphics[keepaspectratio=true,width=150mm]{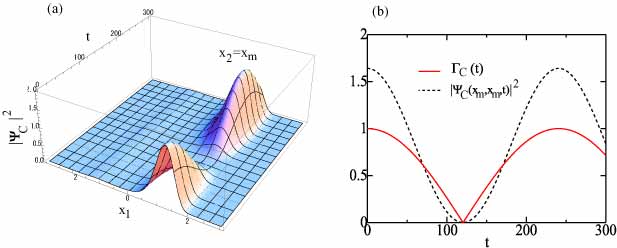}
\end{center}
\caption{
(Color online) 
(a) 3D plot of $\vert \Psi_C(x_1,x_m,t) \vert^2$ as functions of $x_1$ and $t$ with $x_m=1.23534$.
(b) Time dependence of $\Gamma_C(t)$ (solid curve) 
and $\vert \Psi_C(x_m,x_m,t) \vert^2$ (dashed curve) for the wavepacket C ($g=0.1$).
}
\label{fig10}
\end{figure}

\begin{figure}
\begin{center}
\includegraphics[keepaspectratio=true,width=80mm]{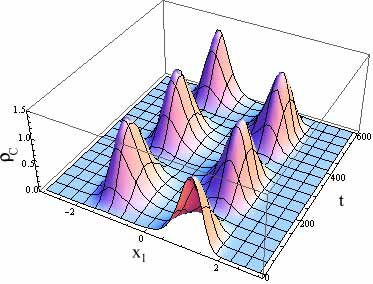}
\end{center}
\caption{
(Color online) 
3D plot of $\rho_{C}(x_1, t)$ as functions of $x_1$ and $t$ for the wavepacket C ($g=0.1$).
}
\label{fig11}
\end{figure}

Figures \ref{fig9}(a)-\ref{fig9}(f) show the time dependence of the magnitude of 
$\vert \Psi_C(x_1,x_2,t) \vert^2$.
Figure \ref{fig9}(a) shows that at $t=0$, the wavepacket given by 
\begin{eqnarray}
\Psi_{C}(x_1,x_2) &=& \frac{1}{\sqrt{2}} 
\left[ \Phi_0(x_1,x_2) +\Phi_1(x_1,x_2) \right],
\label{eq:C3}
\end{eqnarray}
has the maximum magnitude at the $RR$ side of $(x_1, x_2)=(x_m, x_m)$.
We note that with time development, the magnitude of wavepacket
at the initial position at the $RR$ side is decreased
while that at the $LL$ side of $(x_1, x_2)=(-x_m, -x_m)$ is increased. 
At $t=0.5 T$, $ \Psi_C(x_1,x_2,t)$ reduces to the state given by
\begin{eqnarray}
\Psi_{C}^{\bot}(x_1,x_2) &=& \frac{1}{\sqrt{2}} 
\left[ \Phi_0(x_1,x_2) -\Phi_1(x_1,x_2) \right],
\label{eq:C4}
\end{eqnarray}
whose magnitude locates at the $LL$ side of $(x_1, x_2)=(-x_m, -x_m)$, and 
which is the orthogonal state to Eq. (\ref{eq:C3}).
This expresses the tunneling of a particle across the potential barrier at the origin
of $(x_1, x_2)=(0.0, 0.0)$. The wavepacket returns to the initial state at $t=T$.

Dynamics of the wavepacket is studied in more detail. 
We show in Fig. \ref{fig10}(a), the 3D plot
of $\vert \Psi_C(x_1,x_m,t) \vert^2$ as functions of $x_1$ and $t$.
Solid and dashed curves in Fig. \ref{fig10}(b) show time dependences of $\Gamma(t)$ and 
$\vert \Psi_C(x_m,x_m,t) \vert^2$, respectively, which
oscillate with a period of $T=2 \tau=240.63$.

By using Eqs. (\ref{eq:H7a})-(\ref{eq:H7b}) and (\ref{eq:C1}), 
we calculate the density probability of the $x_1$ component
\begin{eqnarray}
\rho_{C}(x_1, t) &=& \int_{-\infty}^{\infty}\: \vert \Psi_C(x_1, x_2, t) \vert^2 \:dx_2, \\
&=&\frac{1}{4} \left[ (2 \cos^2 \theta+1) \:\phi_0(x_1)^2 
+ (2 \sin^2 \theta+1) \:\phi_1(x_1)^2 \right] \nonumber \\
&+& \frac{1}{\sqrt{2}}(\cos \theta+\sin \theta)\: \phi_0(x_1)\phi_1(x_1)\:\cos \Omega_1 t. 
\end{eqnarray}
The time-dependent expectation value of $\langle x_1 \rangle$ is expressed by
\begin{eqnarray}
\langle x_1 \rangle &=& \int_{-\infty}^{\infty} \:\rho_C(x_1, t) \:x_1 \:dx_1, \\
&=& \frac{\gamma}{\sqrt{2}}\:(\cos \theta+\sin \theta)\:\cos \Omega_1 t,
\end{eqnarray}
where $\gamma$ is given by Eq. (\ref{eq:H5}).
Figure \ref{fig11} shows the 3D plot of $\rho_{C}(x_1, t)$.
Similar analysis may be made for the component $x_2$.
If we read $x_1 \rightarrow x_2$ in Fig. \ref{fig11}, it expresses the density probability
for the $x_2$ component.

A comparison between Figs. \ref{fig6}(a) and \ref{fig9}(a) indicates that
$\Psi_C(x_1, x_2, 0.0)$ is initially similar to $\Psi_A(x_1, x_2, 0.0)$, both of which
have appreciable magnitudes at the RR site. Nevertheless, their time development is quite different:
{\it e.g.} $\Psi_C(x_1, x_2, 0.2T) \neq \Psi_A(x_1, x_2, 0.2 T)$. 

\subsubsection{Wavepacket D: $a_0=a_1=a_2=a_3=1/2$}
\begin{figure}
\begin{center}
\includegraphics[keepaspectratio=true,width=150mm]{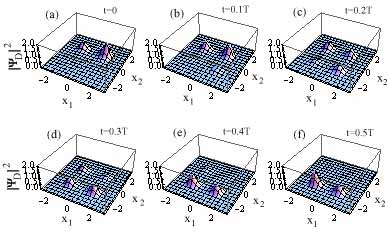}
\end{center}
\caption{
(Color online) 
Time-dependent magnitudes of $\vert \Psi_D(x_1,x_2,t) \vert^2$ 
for (a) $t=0$, (b) $t=0.1 T$, (c) $t=0.2 T$, (d) $t= 0.3 T$, (e) $t=0.4 T$ and (f) $t= 0.5 T$
for the wavepacket D [Eq. (\ref{eq:D1})] with $g=0.1$ where $T=242.32$.
}
\label{fig12}
\end{figure}

\begin{figure}
\begin{center}
\includegraphics[keepaspectratio=true,width=150mm]{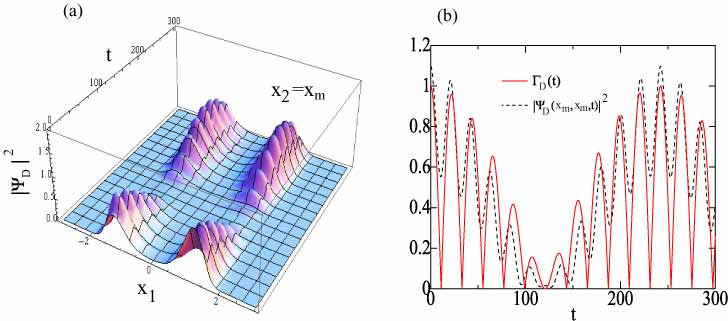}
\end{center}
\caption{
(Color online) 
(a) 3D plot of $\vert \Psi_D(x_1,x_m,t) \vert^2$ 
as functions of $x_1$ and $t$ with $x_m=1.23534$.
(b) Time dependence of $\Gamma_D(t)$ (solid curve) 
and $\vert \Psi_D(x_m,x_m,t) \vert^2$ (dashed curve) for the wavepacket D ($g=0.1$).
}
\label{fig13}
\end{figure}

With $a_0=a_1=a_2=a_3=1/2$, Eq. (\ref{eq:G1}) yields the wavepacket given by
\begin{eqnarray}
\Psi_{D}(x_1,x_2, t) &=& 
\frac{1}{2} \sum_{\nu=0}^3 \:\Phi_{\nu}(x_1,x_2)\:e^{-i E_{\nu} t/\hbar},
\label{eq:D1}
\end{eqnarray}
which leads to the correlation function
\begin{eqnarray}
\Gamma_D(t) &=& \frac{1}{4}\vert 1+ e^{-i \Omega_1 t}
+ e^{-i \Omega_2 t}+ e^{-i \Omega_3 t} \vert,
\label{eq:D2}
\end{eqnarray}
with $\Omega_1=0.02611$, $\Omega_2=0.28522$ and $\Omega_3=0.31134$.
The time-dependent $\vert \Psi_D(x_1,x_2,t) \vert^2$ 
from $t=0$ to $t=0.5 T$ are shown in Figs. \ref{fig12}(a)-\ref{fig12}(f) where $T=242.32$ (below).
In order to scrutinize its behavior, we show in Fig. \ref{fig13}(a), the 3D plot
of $\vert \Psi_D(x_1,x_m,t) \vert^2$ as functions of $x_1$ and $t$.
The dashed curve in Fig. \ref{fig13}(b) expresses 
$\vert \Psi_D(x_m,x_m,t) \vert^2$ whereas the solid curve shows $C_D(t)$
which is expressed as a superposition of three oscillations with frequencies of
$\Omega_1$, $\Omega_2$ and $\Omega_3$.
Both $C_D(t)$ and $\vert \Psi_D(x_m,x_m,t) \vert^2$ show rapid and complicated oscillations
with zeros of $C_D(t)$ at $t=11.01 \:(2k+1)$ with $k=0, 1, \cdots$.
We obtain 
\begin{eqnarray}
T &=& 242.32 \simeq \frac{2 \pi}{E_1-E_0} 
=240.63, \\
\tau &=& 11.01 \simeq \frac{\pi}{E_3-E_1}=11.02.
\end{eqnarray}
The tunneling period $T$ is mainly determined by a energy gap between $E_0$ and $E_1$, 
while a small $\tau$ originates from a large energy gap between $E_1$ and $E_3$.

\section{Discussion}
\subsection{Comparison among results of four wavepackets A, B, C and D}
It has been pointed out \cite{Giovannetti03,Giovannetti03b}  
that the entanglement enhances the speed of evolution in certain quantum state 
as measured by the time speed to reach an orthogonal state.
The orthogonality time $\tau$ is shown to be given by \cite{Giovannetti03,Giovannetti03b}
\begin{eqnarray}
\tau &\geq& \tau_{\min} 
\equiv \max \left( \frac{\pi \hbar}{2 E}, \;\;\frac{\pi \hbar}{2 \Delta E} \right),
\label{eq:J1}
\end{eqnarray}
where $E$ and $\Delta E$ signify expectation and root-mean-square values,
respectively, of the energy relative to $E_0$,
\begin{eqnarray}
E 
&=& \sum_{\nu}\: \vert a_{\nu}\vert^2 \:(E_{\nu}-E_0), 
\label{eq:J2}\\
\Delta E 
&=& \sqrt{ \sum_{\nu}\: \vert a_{\nu}\vert^2 \:(E_{\nu}-E_0)^2-E^2 }.
\label{eq:J3}
\end{eqnarray}
Equations (\ref{eq:J1})-(\ref{eq:J3}) show that the minimum orthogonality time $\tau_{min}$ 
depends on the distribution of eigenvalues and the expansion coefficient of wavepackets.
Applying Eqs. (\ref{eq:J1})-(\ref{eq:J3}) to our DW model
in Eqs. (\ref{eq:H6a})-(\ref{eq:H6b}), we have evaluated $E$, $\Delta E$ and 
$\tau_{min}$ whose results are summarized in the Table 2.
We note that $\tau_{min}$ is determined by $E$ ($< \Delta E$) for the wavepacket C, 
while it is determined by $\Delta E$ ($< E$) for wavepackets A and D 
($E=\Delta E$ for the wavepacket B).

The tunneling period $T$ and the orthogonality time $\tau$
for four wavepackets A, B, C, and D calculated in the preceding section 
are summarized in Table 2.
It is shown that $\tau$ in the four wavepackets 
are in agreement with results of $\tau \geq \tau_{min}$ evaluated by Eqs. (\ref{eq:J1})-(\ref{eq:J3}).
$\tau$ of the entangle wavepacket B is smaller than that of 
the non-entangled wavepacket A ($g=0.0$), which is consistent with 
an enhancement of $\tau$ by entanglement
in uncoupled qubits \cite{Giovannetti03,Giovannetti03b}.

\begin{center}
\begin{tabular}[t]{|c|c|c|c|c|c|c|c| }
\hline
wavepacket & $g$ & $T$ & $\tau$ & $E$ &$\;\;\Delta E \;\;$ & $\tau_{min}$ & $\;\; C \;\;$ \\ 
\hline \hline
$\;\;\;$ A $\;\;\;$ & $\;\;0.0 \;\; $  &  $\;\;\;\; 72.81 \;\;\;\;$  &  $\;\; 36.40 \;\;$ 
& $\;\;\; 0.0863 \;\;\;$ & $\;\;\; 0.0610 \;\;\;$ & $\;\;\; 25.74 \;\;\;$ & 0.0 \\
\hline
B & $ 0.0 $  &  36.40  &  18.20  & 0.0863& 0.0863 & 18.20 & 1.0 \\
\hline
C & $ 0.1 $ & 240.63 &  120.32 & 0.0131 &  0.1213 & 119.91 & $\;\;0.0839 \;\;$ \\
\hline
D & $ 0.1 $ &  242.32 &  $\;\;11.02 \;\;$ & 0.1557 & 0.1432 & 10.97 & 0.2772 \\
\hline
\end{tabular}
\end{center}
{\it Table 2} The tunneling period $T$ [Eq. (\ref{eq:E1})], 
the orthogonality time $\tau$ [Eq. (\ref{eq:E2})], 
the expectation value of the energy $E$ [Eq. (\ref{eq:J2})], 
the root-mean-square value $\Delta E$ [Eq. (\ref{eq:J3})], 
the minimum orthogonality time $\tau_{min}$ [Eq. (\ref{eq:J1})], 
and the concurrence $C$ [Eq. (\ref{eq:K4b})]
in four wavepackets A, B, C and D with couplings $g$ (see Table 1). 

\subsection{Calculation of the concurrence}
In order to examine the relation between $\tau_{min}$ and the entanglement,
we have calculated the concurrence which is one of typical measures 
expressing the degree of entanglement.
Substituting Eqs. (\ref{eq:H7a})-(\ref{eq:H7b}) to Eq. (\ref{eq:G1}) with $t=0$, 
we obtain
\begin{eqnarray}
\vert \Psi \rangle &=& c_{00} \vert 0\;0 \rangle+ c_{01} \vert 0\;1 \rangle
+ c_{10} \vert 1\;0 \rangle + c_{11} \vert 1\;1 \rangle,
\label{eq:K1}
\end{eqnarray}
with
\begin{eqnarray}
c_{00} &=& a_0 \:\cos \theta-a_3 \:\sin \theta, 
\;\;c_{01} = \frac{1}{\sqrt{2}}(a_1-a_2), \nonumber \\
c_{10} &=& \frac{1}{\sqrt{2}}(a_1+a_2),
\;\; c_{11} = c_0 \:\sin \theta+a_3 \:\cos \theta,
\label{eq:K2}
\end{eqnarray}
where $\vert k \;\ell \rangle=\phi_k(x_1) \phi_{\ell}(x_2)$ with $k, \ell=0,1$.
The concurrence $C$ of the state $\vert \Psi \rangle$ given by Eq. (\ref{eq:K1}) is defined by
\cite{Wootters01}
\begin{eqnarray}
C &=& 2 \:\vert c_{00} c_{11}- c_{01} c_{10} \vert. 
\label{eq:K4}
\end{eqnarray}
The state given by Eq. (\ref{eq:K1}) becomes factorizable if and only if the
relation: $c_{00} c_{11}- c_{01} c_{10} =0$ holds.
Substituting Eq. (\ref{eq:K2}) into Eq. (\ref{eq:K4}), we obtain the concurrence 
\begin{eqnarray}
C &=& \vert (a_0^2-a_3^2) \sin 2 \theta+ 2 a_0 a_3 \cos 2\theta- a_1^2+a_2^2 \vert.
\label{eq:K4b}
\end{eqnarray}
By using adopted coefficients in Table 1, we obtain the concurrence for the four wavepackets
\begin{eqnarray}
C_A &=& \frac{1}{2} \;\vert 1- \cos 2 \theta \vert \hspace{1cm}\mbox{(wavepacket A)}, 
\label{eq:K5a}\\
C_B &=& \vert \cos 2 \theta \vert \hspace{2cm}\mbox{(wavepacket B)}, 
\label{eq:K5b}\\
C_C &=& \frac{1}{2} \;\vert 1- \sin{2 \theta} \vert \hspace{1cm}\mbox{(wavepacket C)}, 
\label{eq:K5c}\\
C_D &=& \frac{1}{2} \; \vert \cos{2 \theta} \vert \hspace{1.5cm}\mbox{(wavepacket D)},
\label{eq:K5d}
\end{eqnarray}
which lead to $C_A=0.0$, $C_B=1.0$ for $g=0.0$
and to $C_C=0.0839$ and $C_D=0.2772$ for $g=0.1$ (see Table 2).

\begin{figure}
\begin{center}
\includegraphics[keepaspectratio=true,width=120mm]{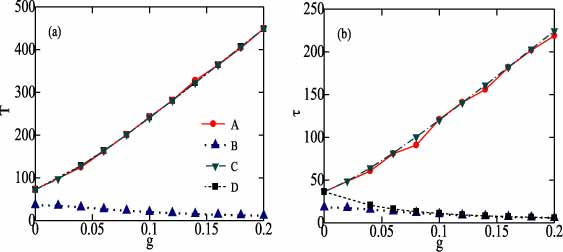}
\end{center}
\caption{
(Color online) 
The $g$ dependence of (a) the tunneling period $T$ and (b) the orthogonality time $\tau$
for wavepacket A (circles), B (triangles), C (inverted triangles) and D (squares).
Results of $T$ for A, C and D are almost identical in (a).
}
\label{fig14}
\end{figure}

\subsection{The $g$ dependence of $T$, $\tau$, $\tau_{min}$ and $C$}
So far calculations are reported only for wavepackets A and B with $g=0.0$ 
and for wavepackets C and D with $g=0.1$.
We have calculated $T$, $\tau$, $\tau_{min}$ and $C$, by changing $g$ 
in a range of $0 \leq g < 0.2$ for four wavepackets A, B, C and D 
whose expansion coefficients $a_{\nu}$ ($\nu=0-3$) are given in Table 1.
For wavepackets B and C consisting of two terms, it is possible 
to exactly calculate the tunneling period and the orthogonality time
with the use of Eq. (\ref{eq:E4}).
However, for wavepackets A and D with more than three terms, 
numerical methods are required for calculations of $T$ and $\tau$. 
Calculated $T$ and $\tau$ are plotted in Figs. \ref{fig14}(a) and \ref{fig14}(b), respectively. 
Our calculations show that $T$ and $\tau$ for the four wavepackets are given by
\begin{eqnarray}
T_{A} &\simeq & T_C = \frac{2 \pi}{E_1-E_0}, 
\;\;\;T_B = \frac{2 \pi}{E_3-E_0}, 
\:\:\:T_D \simeq  \frac{2 \pi}{E_1-E_0}, 
\label{eq:K6}\\
\tau_{A} &\simeq & \tau_C = \frac{\pi}{E_1-E_0}, 
\;\;\;\tau_B = \frac{\pi}{E_3-E_0}, 
\;\;\;\tau_D \simeq \frac{\pi}{E_3-E_1},
\label{eq:K7}
\end{eqnarray}
where $E_{\nu}$ ($\nu=0 - 3$) are $g$ dependent [Eqs. (\ref{eq:H6a})-(\ref{eq:H6b})].
Figure \ref{fig14}(a) shows that
with increasing $g$, the tunneling period is increased for wavepackets A, C and D
while it is decreased for the wavepacket B. This is 
because a gap of $E_1-E_0$ ($E_3-E_0$) is decreased (increased) with increasing $g$ (Fig. 3).
Due to the similar reason, the orthogonality time for wavepackets A and C are increased 
with increasing $g$ whereas it is decreased for wavepackets B and D, as shown in Fig. \ref{fig14}(b).

\begin{figure}
\begin{center}
\includegraphics[keepaspectratio=true,width=120mm]{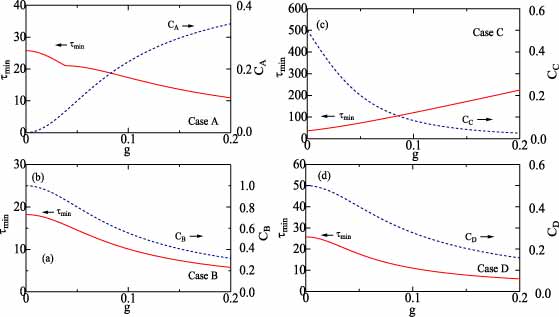}
\end{center}
\caption{
(Color online) 
The minimum orthogonality time $\tau_{min}$ (solid curves) and the concurrence $C$ (dashed curves)
as a function of the interaction $g$ for the wavepackets (a) A, (b) B, (c) C and (d) D,
left and right ordinates being for $\tau_{min}$ and $C$, respectively.
}
\label{fig15}
\end{figure}
 
\begin{figure}
\begin{center}
\includegraphics[keepaspectratio=true,width=115mm]{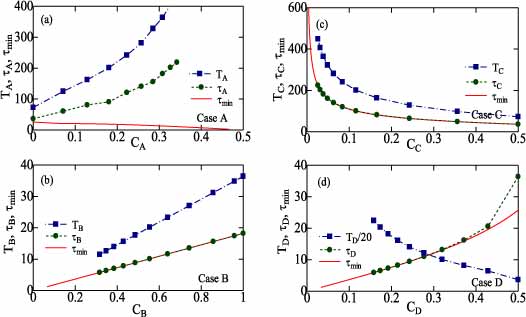}
\end{center}
\caption{
(Color online) 
$T$ (chain curves), $\tau$ (dashed curve) and $\tau_{min}$ (solid curves) 
as a function of the concurrence $C$ for wavepackets (a) A, (b) B, (c) C and (d) D,
$T$ in (d) being divided by a factor of 20.
}
\label{fig16}
\end{figure}
 
$g$ dependences of $\tau_{min}$ and $C$ calculated with the use of Eqs. (\ref{eq:J1})-(\ref{eq:J3}) 
and Eqs. (\ref{eq:K5a})-(\ref{eq:K5d}) for the four wavepackets are shown 
in Figs. \ref{fig15}(a)-\ref{fig15}(d).
Figure \ref{fig15}(a) shows that with increasing $g$ for the wavepacket A, the concurrence
is increased from a vanishing value of $C=0.0$ at $g=0.0$ while
$\tau_{min}$ is decreased: a kink of $\tau_{min}$ at $g=0.0366$ is due to a crossover of 
$\pi \hbar/2 E= \pi \hbar/2 \Delta E$ in Eq. (\ref{eq:J1}).
Figure \ref{fig15}(b) shows that for the wavepacket B, an increase in $g$ induces a decrease in $\tau_{min}$ 
and $C$, the latter being decreased from the maximum concurrence of $C=1.0$ at $g=0.0$.
For the wavepacket C, $\tau_{min}$ is increased but $C $ is decreased with increasing $g$,
as shown in Fig. \ref{fig15}(c).
Figure \ref{fig15}(d) shows that both $\tau_{min}$ and $C$ are decreased 
with increasing $g$ for the wavepacket D.

Figures \ref{fig16}(a)-\ref{fig16}(d) show $T$, $\tau$ and $\tau_{min}$ as 
a function of the concurrence $C$ for the four wavepackets.
It is shown that for a larger $C$, $T$ is larger in wavepackets A and B, while
it is smaller in wavepackets C and D.
We note that for a larger concurrence, $\tau$ is larger for wavepackets A, B and D,  
but it is smaller for the wavepacket C.
For a larger $\tau$, $T$ is larger in wavepackets A, B and C, but it is not the case 
for the wavepacket D. This fact imposes a question whether the evolution time may be measured
by $T$ or $\tau$. Furthermore, the speed of quantum evolution measured by either $\tau$ or $T$ 
is not necessarily increased when $C$ is increased. This is in contrast 
with Refs. \cite{Giovannetti03,Giovannetti03b}
which claimed that the speed of a development of quantum state is improved by the entanglement.
We also note that $\tau_{min}$ given by Eq. (\ref{eq:J1}) provides us with fairly good estimates 
for lower limits of $\tau$ for wavepackets B, C and D. However, it does not for the wavepacket A.
In order to clarify the point, we show
the correlation functions $\Gamma_A(t)$ with $g=0.0$, $0.1$ and $0.2$ 
for the wavepacket A in Figs. \ref{fig5}(a), \ref{fig5}(b) and \ref{fig5}(c), respectively.
$\Gamma_A(t)$ with $g=0.1$ and $0.2$ more rapidly oscillates than that with $g=0.0$.
We obtain $( \tau, \tau_{min})=(36.40,\;25.74)$, $(121.00,\;17.28)$ and $(218.77,\;12.30)$
for $g=0.0$, 0.1 and 0.2, respectively.
With increasing $g$, $\tau$ is increased because of a narrowed energy gap of $E_1-E_0$ 
in Eq. (\ref{eq:K7}), whereas $\tau_{min}$ is decreased by a high-energy contribution 
of $E_3-E_0$ in Eqs. (\ref{eq:J1})-(\ref{eq:J3}).
Although the relation: $\tau_{min} \leq \tau$ is actually held, 
the difference between $\tau$ and $\tau_{min}$ is significant with increasing $g$ 
for the wavepacket A, where $\tau_{min}$ given by Eqs. (\ref{eq:J1})-(\ref{eq:J3}) is not 
a good estimate of the lower bound of $\tau$ determined by Eq. (\ref{eq:E2}).

\section{Conclusion}
With the use of an exactly solvable coupled DW system described by Razavy's potential \cite{Razavy80}, 
we have studied the dynamics of four wavepackets A, B, C and D (Table 1). 
Our calculations of tunneling period $T$ and the orthogonality time $\tau$ yield the followings:

\noindent
(1) Although the relation: $T = 2 \tau$ holds for wavepackets B and C including two terms,
it is not the case in general. In particular for the wavepacket D, $T$ is increased
but $\tau$ is decreased with increasing $g$ (Fig. \ref{fig14}), and

\noindent
(2) $g$ dependences of $T$ and $\tau$ considerably depend on a kind of adopted wavepackets
(Fig. \ref{fig14}), and they are increased or decreased with increasing the concurrence,
depending on an initial wavepacket (Figs. \ref{fig15} and \ref{fig16}).

\noindent
A query arises from the item (1) whether the speed of a quantum evolution
may be measured by $T$ or $\tau$, although it is commonly evaluated by $\tau$
\cite{Giovannetti03,Giovannetti03b,Batle05,Curilef06,Zander13}.
The item (2) implies that even if the evolution speed is measured by either $\tau$
or $T$, it is not necessarily increased by the entanglement. 
This is in contrast with Refs. \cite{Giovannetti03,Giovannetti03b,Batle05,Curilef06,Zander13} 
which pointed out an enhancement of the evolution speed by the entanglement in TL models.
The difference between their results and ours arises from the fact that
the coupled DW model has much freedom than the TL model: the latter is a simplified
model of the former.
It would be interesting to experimentally observed the time-dependent magnitude of
$\vert \Psi(x_1, x_2, t) \vert^2$, which might be possible with advanced recent technology.
In the present study, we do not take into account environmental effects which are expected
to play important roles in real DW systems. An inclusion of dissipative effects is left as
our future subject.

\begin{acknowledgments}
This work is partly supported by a Grant-in-Aid for Scientific Research from 
Ministry of Education, Culture, Sports, Science and Technology of Japan.  
\end{acknowledgments}


\end{document}